\documentclass[prl,twocolumn,aps,amsmath,nofootinbib,superscriptaddress]{revtex4}
\usepackage{graphicx}
\usepackage{bm}
\usepackage{epsfig}

\newif\ifpdf
\ifx\pdfoutput\undefined
\pdffalse 
\else
\pdfoutput=1 
\pdftrue
\fi

\def\Slash#1{{#1\!\!\!\slash}}

\def\Bslash{B\!\!\!\!\slash}
\def\Dslash{D\!\!\!\!\slash}

\def\Mslash{M\!\!\!\!\!\slash}

\def\nslash{n\!\!\!\slash}
\def\bnslash{\bar n\!\!\!\slash}
\def\pslash{p\!\!\!\slash}

\def\OMIT#1{}

\newcommand{\nn}{\nonumber} 

\newcommand{\bn}{{\bar n}}
\newcommand{\bea}{\begin{eqnarray}}
\newcommand{\eea}{\end{eqnarray}}

\newcommand{\bnP}{\bar {\cal P}}

\newcommand{\cP}{{\cal P}}

\newcommand{\mcdot}{\!\cdot\!}

\newcommand{\LQCD}{{\Lambda}}

\newcommand{\SCETa}{\mbox{${\rm SCET}_{\rm I}$ }}
\newcommand{\SCETb}{\mbox{${\rm SCET}_{\rm II}$ }}

\newcommand{\DgppPl}{\,\overleftarrow D{}_{\!c\,\alpha}^{\perp}}
\newcommand{\DgppPr}{\,\overrightarrow D{}_{\!c\,\alpha}^{\perp}}

\begin{document}
\ifpdf
\DeclareGraphicsExtensions{.pdf, .jpg}
\else
\DeclareGraphicsExtensions{.eps, .jpg}
\fi


\preprint{ \vbox{\hbox{UCSD/PTH 02-27} \hbox{hep-ph/0211069}  }}

\title{
Factorization and Endpoint Singularities in Heavy-to-Light decays}

\author{Christian W.~Bauer}
\affiliation{Department of Physics, University of California at San Diego,
        La Jolla, CA 92093}
\author{Dan Pirjol}
\affiliation{Department of Physics and Astronomy, The Johns Hopkins University,
        Baltimore, MD 21218}
\author{Iain W. Stewart\vspace{0.4cm}}
\affiliation{Institute for Nuclear Theory,  University of Washington, Seattle, 
        WA 98195 
        \vspace{0.3cm}
}

\begin{abstract}
  
  We prove a factorization theorem for heavy-to-light form factors. Our result
  differs in several important ways from previous proposals. A proper separation
  of scales gives hard kernels that are free of endpoint singularities. A
  general procedure is described for including soft effects usually associated
  with the tail of wavefunctions in hard exclusive processes. We give an
  operator formulation of these soft effects using the soft-collinear effective
  theory, and show that they appear at the same order in the power counting as
  the hard spectator contribution.

\end{abstract}

\maketitle

Exclusive hadronic form factors simplify dramatically at momentum transfers much
larger than hadronic scales, $Q^2\gg\Lambda^2$. Typically, they factor into
non-perturbative light cone wavefunctions $\phi_{a,b}$ for mesons $a$ and $b$,
convoluted with a calculable hard scattering kernel $T$~\cite{bl},
\begin{eqnarray} \label{F1}
 F(Q^2) = \frac{f_a f_b}{Q^2} \int\:\!\!\!\! dx\, dy\: 
  T(x,y,\mu) \phi_a(x,\mu) \phi_b(y,\mu)+ \ldots
\end{eqnarray}
Here $f_i$ are meson decay constants, the hard scattering kernel $T(x,y)$ is
calculated perturbatively in an expansion in $\alpha_s$, and the ellipses
denotes terms suppressed by additional powers of $1/Q$. For example, the
electromagnetic form factor of a pion has $a=b=\pi$ and at $\mu=Q$~\cite{bl}
$T(x,y) = 8 \pi \alpha_s(Q)/(9x y)$. For Eq.~(\ref{F1}) to be well defined it is
sufficient that $\phi_i(x)\stackrel{x\to 0}{\sim} x^n$, $\phi_i(x)\stackrel{x\to
  1}{\sim} (1-x)^m$ with any $n,m>0$. A linear falloff is sometimes assumed, but
we will not use this assumption.

Beyond leading order (LO) in $1/Q$ issues arise. There are soft contributions to
the form factor, which arise from configurations where a single quark carries
most of the meson momentum and leaves $p^\mu\sim \LQCD$ for the remaining
constituents~\cite{QCDsumsoft}, and these  have been estimated using QCD
sum rules. Furthermore, power suppressed hard exchange contributions tend to
give contributions diverging as $\int\! dx/x$. Examples include $1/Q$
corrections to the pion form factor, $1/m_b$ corrections in $B\to \pi\pi,K\pi$
decays, and one-gluon exchange for heavy-to-light form factors~\cite{sing}.

The soft-collinear effective theory (SCET)~\cite{bfps,bfl,cbis,bps2}, reproduces
the factorization in Eq.~(\ref{F1})~\cite{bfprs}, and provides a framework to
analyze power corrections based solely on QCD.  This theory consists of
collinear fields interacting with soft or ultrasoft (usoft) degrees of freedom.
The fields are categorized by the scaling of their momenta: collinear $p_c =
(p_c^+, p_c^-, p_c^\perp)=(n\cdot p_c, \bn\cdot p_c, p_c^\perp) \sim
Q(\lambda^2, 1, \lambda)$, soft $p_s^\mu \sim Q\lambda$ and usoft $p_{us}^\mu
\sim Q\lambda^2$, where $n^2=\bn^2=0$, $n\mcdot\bn=2$, and $\lambda \ll 1$ is
the expansion parameter.

In this paper we show how SCET can be used to understand factorization and
soft-endpoint contributions in heavy-to-light form factors for decays such as
$B\to \pi\ell\nu$, $B\to K^*e^+e^-$ and $B\to \rho\gamma$, building
on~\cite{bfps}. Here the large scales are $Q=\{m_b,E\}$, where the final meson
has $E=m_B/2-q^2/(2m_B)$. Several ideas are developed, which we summarize as an
outline.\\[-8pt]

\hspace{-0.9cm}\begin{minipage}{3.6in}
\begin{itemize}
 \item We prove a factorization formula for heavy-to-light decays involving
 the LO light-cone wavefunctions, a jet function, plus a reduced set
 of non-perturbative matrix elements which obey form factor relations. 
\\[-18pt]
 \item Calculable kernels are free of divergences. 
 Endpoint singularities are fake and arise from improperly matching onto
 $T(x,y)$. They appear in non-factorizable operators
 and can be parameterized without invoking
 suppression from Sudakov effects.  
 \\[-18pt]
 \item A single collinear meson state can be used to categorize all
 contributions. Soft effects associated with the tail of wavefunctions are
 described by matrix elements of operators with a definite power counting.  The
 categories ``factorizable" and ``non-factorizable" are more accurate than 
 ``hard" and ``soft'' contributions.  \\[-18pt]
 \item There are two perturbative scales in the problem: $Q$ and $\mu_{\rm 0}
 \simeq \sqrt{Q \Lambda}$. We separate these scales by matching in two stages,
 onto a \SCETa at $\mu =Q$, and onto a \SCETb at $\mu = \mu_{\rm 0}$.  \\[-18pt]
\item The LO result for heavy-to-light decays comes from power suppressed 
operators in \SCETa\!, which match onto LO operators in \SCETb.
\end{itemize}
\vspace{-0.2cm}
\end{minipage}
Our procedure is quite general and similar analyses apply to other exclusive
processes.

\begin{figure}[!b]
\vskip-0.3cm
 \centerline{
  \mbox{\epsfxsize=3.3truecm \hbox{\epsfbox{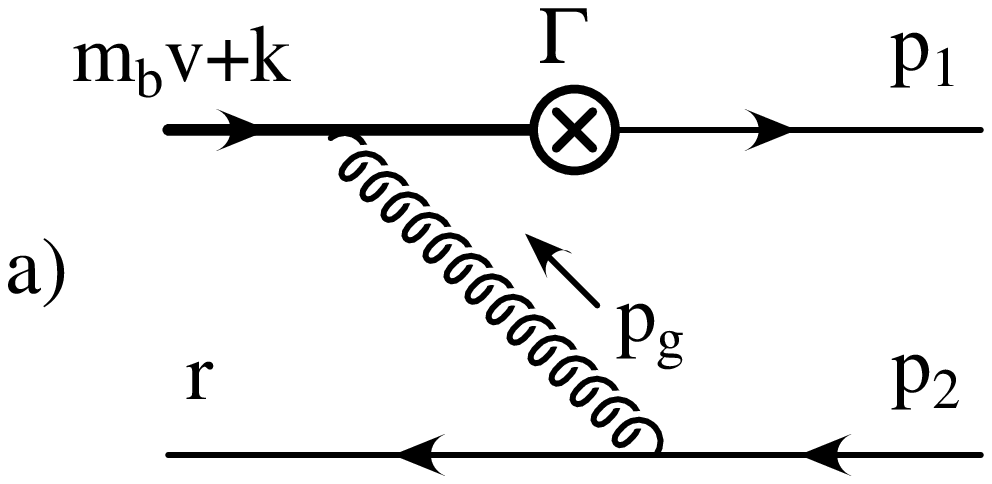}} }\hspace{0.4cm}
  \mbox{\epsfxsize=3.3truecm \hbox{\epsfbox{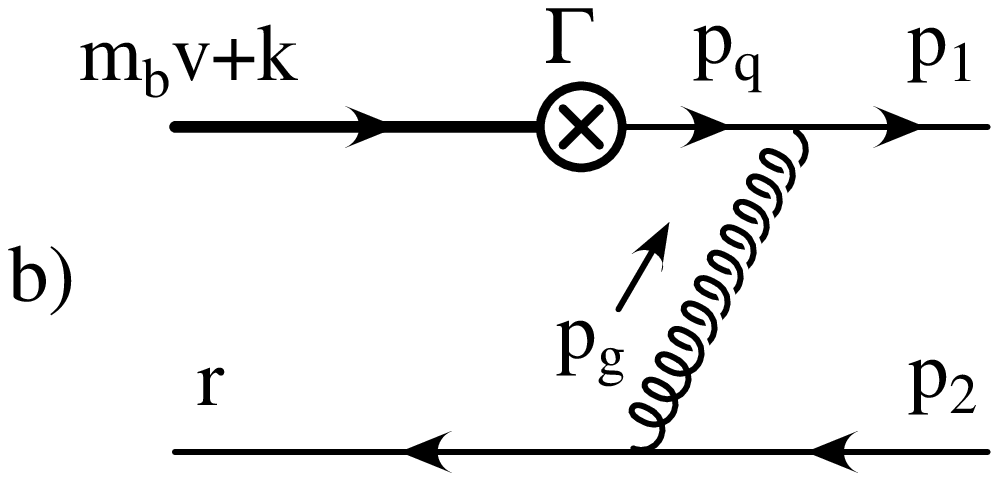}} }
  } 
\vskip-0.3cm
\caption[1]{Tree level QCD graphs for heavy-to-light decays with one 
perturbative gluon. Note that $p_g^2\sim Q\Lambda$.}
\label{fig_qcd} 
\vskip-0.5cm
\end{figure}

To understand the origin of the endpoint divergences, we consider the spectator
interaction for heavy-to-light decays at ${\cal O}(g^2)$ in Fig.\ref{fig_qcd}.
Taking $p_{1,2}$ collinear and $k,r$ ultrasoft, the $\lambda$-expansion of these
graphs gives $i{\cal A}= g^2\bar u_n(p_1) X T^A u_v(p_b) \bar v(r) V T^A
v_n(p_2)\, /P_g$ with
\begin{eqnarray} \label{Wab}
 (X\otimes V)^{(a)}  \!\!&=&\!\!  
   \frac{\Gamma \otimes \bnslash}{\bn\mcdot p_2}
   + \frac{\Gamma \nslash \gamma^\perp_\mu \otimes \gamma_\perp^\mu}
   {2m_b} +\ldots \,, \\
 (X\otimes V)^{(b)} \!\!&=&\!\!  \Big\{
 \frac{\pslash_1^\perp \gamma_\perp^\mu \bn\mcdot p }{P_q\,\bn\mcdot p_1} 
  + \frac{\gamma_\perp^\mu}{P_q} \Big( \frac{n\mcdot p \bnslash}{2}+\Slash{r} 
 \Big)
 \Big\} \Gamma \otimes \gamma^\perp_\mu\nn\\
 &&\hspace{-2cm}
   - \frac{2\bn\mcdot p\, \Gamma\otimes \pslash_2^\perp}{P_q \bn\mcdot p_2} 
   + \frac{\bn\mcdot p\, \pslash_1^\perp\gamma_\perp^\mu \Gamma \otimes 
   \gamma^\perp_\mu \pslash_2^\perp\bnslash}
    {2 P_q\, \bn\mcdot p_1 \bn\mcdot p_2} 
   \bigg\} +\ldots  ,\nn
\end{eqnarray}
where $P_g \!=\! \bn \cdot p_2\, n \cdot r$, $P_q \!=\! \bn \cdot p\, n \cdot
r\!+\! p^2$ and $p=p_1\!-\!p_2$. Eq.~(\ref{Wab}) agrees with Ref.~\cite{bf}. The
$\ldots$ denotes terms $\propto p_\perp$. If one interprets Eq.~(\ref{Wab})
using Eq.~(\ref{F1}) with $a=M$, $b=B$, it is tempting to extract $T(x,y)$
setting $p_i^\perp = 0$, $(p_1-p_2)^2 = 0$, $\bn\cdot p_2=2 x E$, $n\cdot r=y$.
However the result includes terms $\propto 1/x^2$ or $1/y^2$ leading to singular
integrals.  Note that in full QCD there are no singularities since they are
regulated by momenta of order $\Lambda$.

Several proposals have been made for dealing with these divergences. One
approach regulates these singularities by introducing transverse parton momenta
and including Sudakov form factors~\cite{Kurimoto,NagashimaLi}. However this
proposal does not include all the non-perturbative contributions, or deal with
the possibility that Sudakov suppression may not be large at $m_b \approx 5$
GeV.  In~\cite{bf} it was shown that at LO these divergences can be reabsorbed
into ``soft'' form factors which satisfy form factor
relations~\cite{charles,bfps}.  However, this analysis was only performed to
order $\alpha_s$. Furthermore neither a rigorous field theoretic definition for
these soft contributions exists, nor does a first principle derivation of their
power counting.

To fully understand these issues requires a factorization formula with the
generality to account for non-perturbative contributions. In this paper we
prove that at leading order in $1/Q$ and all orders in $\alpha_s$ a generic
heavy-to-light form factor $F$ can be split into factorizable and
non-factorizable components $F = f^{\rm F}(Q) + f^{\rm NF}(Q)$ where
\begin{eqnarray}\label{fF}
f^{F}(Q) &=& N_0 \int_0^1\!\!\!\! dz\! 
    \int_0^1\!\!\!\! dx\! \int_0^\infty\!\!\!\!\! dr_+ \,
    T(z,Q,\mu_{\rm 0})  
    \\
 && \quad \times J(z,x,r_+,Q,\mu_{\rm 0},\mu) \phi_M(x,\mu) \phi_B(r_+,\mu) \,,
  \nn\\
  f^{\rm NF}(Q) &=&C_k(Q,\mu)\: \zeta_k(Q,\mu) \,,
\label{fNF}
\end{eqnarray}
and $N_0=f_B f_\pi m_B/(4 E^2)$.  The hard coefficients $C_k$ and $T$ can be
calculated in an expansion in $\alpha_s(Q)$, the jet function $J$ is dominated
by momenta $p^2\simeq Q\Lambda$ and calculable perturbatively in
$\alpha_s(\sqrt{Q\LQCD})$.  The functions $\phi_M$ and $\phi_B$ are standard
non-perturbative light-cone wave-functions c.f.~\cite{grozin,bf}, where our
$\phi_B$ denotes $\phi_B^+$ or $\phi_B^-$.  Only $\phi_B^+$ appears if $J$ is
calculated at tree level.  Endpoint singularities only arise in matrix elements
which determine the soft, non-perturbative form factors $\zeta_k(Q,\mu)$,
leaving the convolution integrals in the factorizable terms finite. There are
three soft form factors $\zeta_k(Q,\mu)$; one for pseudoscalar, and two for
vector mesons. We show that terms proportional to $\phi_B^-$ can be absorbed
into a redefinition of $\zeta_k(E,\mu)$ at any order in perturbation theory.

\begin{figure}[!t]
 \centerline{
  \mbox{\epsfxsize=5.5truecm \hbox{\epsfbox{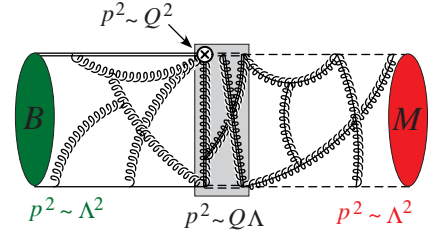}} }
  }
\vskip-0.3cm
\caption[1]{Levels of factorization. The gray area corresponds to gluons in \SCETa which are integrated out in \SCETb.}
\label{fig_FACT} 
\vskip-0.5cm
\end{figure}

Our expression for the heavy to light form factors differs in several important
ways from previous proposals. In the pQCD approach \cite{Kurimoto} possible
non-perturbative soft contributions are dropped with the ex post facto
assumption that they are negligible. Furthermore, their perturbative pieces
contain singular terms in the hard kernels, which are then regulated by
resummations~\cite{Li}. In Ref.~\cite{bf} both soft and non-singular hard
contributions were included. Unlike \cite{Kurimoto} the soft pieces were found
to dominate, due to the fact that the hard terms were suppressed by
$\alpha_s(\sqrt{Q\Lambda})$.  However, their soft and hard definitions do not
clearly avoid double counting.  Furthermore, in order to show that these two
contributions are the same order in $1/m_b$ it was necessary to use assumptions
about the scaling of the tails of the meson wave functions.

In contrast, in our work the $f^F$ and $f^{NF}$ pieces appear from matrix
elements of distinct operators with the same states, avoiding any possibility of
double counting. Furthermore, singular hard scattering kernels do not appear. We
prove a factorization theorem to all orders in perturbation theory. Our result
differs from the proposed formula in Ref.~\cite{bf} because it involves both a
hard kernel $T$ and a jet function $J$, which separate the scales $Q$ and
$\sqrt{Q \Lambda}$.  This separation is necessary if one wants to distinguish
factors of $\alpha_s(Q)$ from $\alpha_s(\sqrt{Q\Lambda})$, or more accurately
resum large logarithms between these scales. Our result differs from
Ref.~\cite{Kurimoto} in that $f^{NF}$ contains non-perturbative matrix elements
of operators with $D_\perp$'s which do not appear in~\cite{Kurimoto}, and our
$f^F$ does not involve $k_\perp$ convolutions.  The operator definitions for the
various pieces in the factorization formula in Eqs.~(\ref{fF},\ref{fNF}) allow
us to rigorously power count the two types of contributions in a model
independent way.  Finally, the form of our result appears to indicate that the
soft and hard contributions may actually be comparable in size, since the
$\alpha_s(\sqrt{Q\Lambda})$ suppression in $J$ could be compensated by a similar
factor in $\zeta_k$.  A complete answer to this question requires a full
resummation of the double Sudakov logarithms, which are known~\cite{ps1} to
appear in both our F and NF contributions. This is left to future work.

To begin, we need a definition of the nonperturbative hadronic states. They can
be defined by any interpolating field which has the right quantum numbers and
significant overlap with the physical state.  For the B we pick the standard
HQET state $|B_v\rangle$~\cite{HQET}, while for the light meson $M$ we pick a
state $| M_n \rangle$ whose interpolating field is built out of two collinear
quarks, and involves all interactions in the LO collinear Lagrangian.  Thus, the
$B$/$M$ states are generated by soft/collinear fields in \SCETb which have
$p_s^2 \sim p_c^2 \sim \Lambda^2$.  Time-ordered products account for
corrections to these states. We do {\em not} define $| M_n \rangle$ with
collinear quarks in \SCETa since here the offshellness is still large.

Eqs.~(\ref{fF},\ref{fNF}) separate the contributions from hard momenta ($p^2
\sim Q^2$), jet momenta ($p^2 \sim Q \LQCD$), and nonperturbative momenta
($p^2\sim \LQCD^2$), as illustrated in Fig.~\ref{fig_FACT}. In
Fig.~\ref{fig_qcd} it is the gluon that connects to the spectator which scales
like a jet momentum. To separate these scales we match QCD onto an intermediate
effective theory \SCETa, valid for $\sqrt{Q\LQCD} < \mu < Q$, which contains
collinear particles with offshellness $p_c^2 \sim Q \LQCD$ and a power counting
in $\lambda = \sqrt{\Lambda/Q}$. Since the collinear particles in \SCETa satisfy
$p_c^2 \sim Q \LQCD$ this theory does not describe the complete $B \to M$
process in QCD.  A second step of matching is required onto \SCETb, containing
collinear particles with offshellness $p_c^2 \sim \LQCD^2$ and power counting in
$\lambda'=\lambda^2=\LQCD/Q$. Wilson coefficients in \SCETa determine $T, C$ of
(\ref{fF}), while those in \SCETb determine $J$. The usoft fields in \SCETa are
identical to soft fields in \SCETb. This two-step procedure provides a simple
and more general method of determining the \SCETb soft-collinear operators
compared to the procedure in Ref.~\cite{bps2}.

\SCETa is defined by its Lagrangian and heavy-to-light currents.  The terms in
the expansion of the collinear Lagrangian we require are ${\cal L}_c = {\cal
  L}_{c}^{(0)} + {\cal L}_{\xi q}^{(1)} + {\cal L}_{\xi\xi}^{(1)} + {\cal
  L}_{cg}^{(1)} + {\cal L}_{\xi q}^{(2a)}+ {\cal L}_{\xi q}^{(2b)}$.  The
superscript denotes the order in $\lambda$ that these terms contribute in the
power counting~\cite{bpspc}. The LO action for collinear quarks and gluons
is~\cite{bfps,bps2}
\begin{eqnarray} \label{L0}
 && {\cal L}^{(0)}_c = \bar \xi_n \left[ i n \mcdot D 
  + i \Slash{D}^c_\perp \frac{1}{i \bn \mcdot D_c}  i \Slash{D}^c_\perp\right]  
  \frac{\bnslash}{2} \xi_n + {\cal L}^{(0)}_{cg} \,,\hspace{0.9cm}
\end{eqnarray}
with $i\bn\mcdot D_c=\bnP\!+\!g\bn\mcdot A_c$, $iD_c^\perp=\cP^\perp\! +
\!gA^\perp_c$, $in\mcdot D=in\mcdot\partial+gn\mcdot A_{us}+gn\mcdot A_c$. The
gluon action ${\cal L}^{(0)}_{cg}$ can be found in Ref.\cite{bps2}.  For the
subleading action we find~\cite{chay,rpi,ps1}
\begin{eqnarray}
&& {\cal L}_{\xi\xi}^{(1)} = \bar \xi_n  i \Slash{D}^{us}_\perp 
  \frac{1}{i \bn \mcdot D_c}  i \Slash{D}^c_\perp \frac{\bnslash}{2} \xi_n
  \mbox{ + h.c.}\,,  \\
&& {\cal L}_{cg}^{(1)} = \frac{2}{g^2} {\rm tr} 
  \Big\{ \big[i {\cal D}^\mu , iD_c^{\perp\nu} \big] 
         \big[i {\cal D}_\mu , iD_{us\,\nu}^\perp \big] \Big\} + {\rm g.f.}\,, 
  \nn
\end{eqnarray}
with ${\cal D}^\mu = n^\mu \bn\mcdot D_c/2 + D_c^{\perp\mu} +\bn^\mu n\mcdot
D/2$ and g.f.~denotes gauge fixing terms.  In our proof the mixed
collinear-usoft Lagrangian ${\cal L}_{\xi q}$ will play a crucial role and was
first considered in \cite{bcdf}.  Using the label operator
formalism~\cite{cbis} we obtain the gauge invariant QCD result:
\begin{eqnarray}\label{Lmix1}
&& {\cal L}^{(1)}_{\xi q} = ig\: \bar\xi_n \: \frac{1}{i\bn\mcdot D_c}\: 
 \Bslash_\perp^c W  q_{us} \mbox{ + h.c.}\,,\\
&& {\cal L}^{(2a)}_{\xi q}  = ig\: \bar\xi_n  \frac{1}{i\bn\mcdot D_c}\: 
  \Mslash  \: W \: q_{us} \mbox{ + h.c.}\,, \nonumber\\\label{Lmix2}
&&{\cal L}^{(2b)}_{\xi q}  =  ig\: \bar\xi_n \frac{\bnslash}{2} 
  i\Dslash_\perp^{\,c} \frac{1}{(i\bn\mcdot D_c)^2}\:  \Bslash_\perp^c W 
  \: q_{us}   \mbox{ + h.c.}\,, \nonumber
\end{eqnarray}
where $ig \Bslash_\perp^c =[i\bn\mcdot D^c,i\Dslash_\perp^{c}]$ and $ig\Mslash
=[i\bn\mcdot D^c,{i\Dslash^{\,us}} +\frac{\bnslash}{2} gn\mcdot A^c]$.  A
possible four quark operator $(\bar\xi_nW T^A \bnslash W^\dagger \xi_n)1/\bnP^2
(\bar\xi_n W T^A\bnslash q_{us})$ has been eliminated using the collinear gluon
equations of motion.  Finally the \SCETa currents we will need
are~\cite{bfps, chay, bps3, bcdf, ps1}
\begin{eqnarray}\label{currents}
 J^{(0)} &=& C_\Gamma(\omega_1) \big(\bar\xi_n W\big)_{\omega_1}\, \Gamma
h_v\\
 J^{(1a)} &=& B_\Gamma^a(\omega_1+\omega_2)\big(\bar \xi_n W\big)_{\omega_1}
\big(W^\dagger i\! 
{\DgppPl} W\big)_{\omega_2}
   \frac{\Gamma^{\alpha}_a}{\bnP^\dagger}   h_v \nn \\
 J^{(1b)} &=& B_\Gamma^b(\omega_1,\omega_2) \big(\bar \xi_n 
W)_{\omega_1} 
  \big(W^\dagger i\! \DgppPr  W\big)_{\omega_2} \frac{\Gamma^{\alpha}_b}{m_b} 
 h_v\,, \nn 
\end{eqnarray}
where we sum over $\omega_1,\omega_2$.  Here $J^{(1a,1b)}$ correspond to the
$K_{j}^{(d)}$ of~\cite{ps1}, which are the most general allowed operators at any
order in $\alpha_s$, taking $v_\perp=0$.

In matching onto \SCETb we need two collinear quarks to give non zero overlap
with $|M_n\rangle$, so we only need operators with two collinear quarks in
\SCETa.  For the graphs in \SCETa it is necessary to have a ${\cal L}_{\xi
  q}^{(n)}$ interaction to turn the usoft spectator in the $B$ into a collinear
quark. This is the generic reason that the form factors in the range of $q^2$
considered here, $q^2 \lesssim 10\,{\rm GeV}^2$, are suppressed relative to
their size near $q^2_{\rm max}$. More than one ${\cal L}_{\xi q}^{(n)}$
insertion is forbidden at this order. The relevant time-ordered products are
\begin{eqnarray}
T_0^{\rm F} \!\!&=&\!\! T[J^{(0)},i{\cal L}^{(1)}_{\xi q}
\big]\equiv \int\!\! d^4\!x\, T \big[J^{(0)}(0)\, i{\cal L}^{(1)}_{\xi q}(x) 
\big]
\end{eqnarray}
 as well as
\begin{eqnarray} \label{Tproducts}
T_1^{\rm F} \!\!&=&\!\! T \big[J^{(1a)}, i{\cal L}^{(1)}_{\xi q} \big],\
 \qquad
 T_2^{\rm F} =\! T \big[J^{(1b)}, i{\cal L}^{(1)}_{\xi q} \big] \,, \\
 T_3^{\rm F} \!&=&\! T \big[J^{(0)},i{\cal L}^{(2b)}_{\xi q} \big],
 \qquad\ \: 
 T_4^{\rm NF} =\! T \big[J^{(0)},i{\cal L}^{(2a)}_{\xi q} \big],\nn\\
 T_5^{\rm NF} \!\!&=&\!\! T \big[J^{(0)}, i{\cal L}^{(1)}_{\xi\xi},
  i{\cal L}^{(1)}_{\xi q} \big], \
  \quad\!\!\!\!\!
 T_6^{\rm NF} =\! T \big[J^{(0)}, i{\cal L}^{(1)}_{cg},i{\cal L}^{(1)}_{\xi q} 
  \big] \,. \nn \hspace{0.cm}
\end{eqnarray}
The time-ordered product $T_0^{\rm F}$ is enhanced by one power of $\lambda$ in
\SCETa compared to the other terms, however its matching onto \SCETb does not
give rise to enhanced contributions to form factors. Higher order $T$'s do not
contribute at the order we are working.

To prove the factorization formula given in (\ref{fF},\ref{fNF}), we decouple the
collinear-usoft interaction in the LO Lagrangian ${\cal L}^{(0)}_c$ using the
field redefinitions~\cite{bps2}
\begin{eqnarray}
 \xi_n^{(0)} = Y^\dagger \xi_n\,, \quad 
 A_n^{(0)} = Y^\dagger A_n Y \,, \\
 Y(x) = \mbox{P}\exp\left( ig\int_{-\infty}^x \mbox{d}s\: n\mcdot 
 A_{us}(ns) \right)\,.\nn
\end{eqnarray}
While this introduces a factor of $Y^\dagger$ into the leading current, it only
appears in the combination ${\cal H}_v = [Y^\dagger h_v]$
\begin{eqnarray}
 J^{(0)} &=& C_\Gamma(\omega_1) \big(\bar\xi_n^{(0)} W^{(0)} \big)_{\omega_1}\, 
 \Gamma \, {\cal H}_v \,.
\end{eqnarray}
The situation is similar in ${\cal L}_{\xi q}^{(1)}$ and ${\cal L}_{\xi
q}^{(2b)}$, where usoft fields/interactions now only appear in the
combination ${\cal Q} = [Y^\dagger q_{us}]$.  On the other hand we  have
\begin{eqnarray}
{\cal L}_{\xi\xi}^{(1)} \!\!&=&\!\! 
  \bar \xi_n^{(0)} \big[ Y^\dagger i\Dslash_{us}^\perp Y \big] \frac{1}
  {i \bn \mcdot D_c^{(0)}} i \Dslash_{c,\perp}^{(0)} \frac{\bnslash}{2} 
  \xi_n^{(0)} \mbox{ + h.c.,} \\
{\cal L}^{(2a)}_{\xi q} \!\!&=&\!\! 
  ig\: \bar\xi_n^{(0)}  \frac{1}{i\bn\mcdot D_c^{(0)}}\,
  \big[ Y^\dagger \Mslash \;\, Y \big] 
  W^{(0)}  {\cal Q} \mbox{ + h.c.} \,. \nn
\end{eqnarray}
Thus, the time-ordered products fall into two categories: ``factorizable'',
$T^{\rm F}_{\{0,1,2,3\}}$, in which the usoft interactions all occur in ${\cal
  H}_v$ and ${\cal Q}$ , and ``non-factorizable'', $T^{\rm NF}_{\{4,5,6\}}$,
with an additional $[Y^\dagger D_{us}^\mu Y]$ or $[Y^\dagger \Mslash \;\, Y]$.
It can be clearly seen that there is no double counting when the soft and hard
contributions are defined this way.  The matching onto \SCETb for these two
cases is discussed separately.

For the factorizable terms $T_i^F=T[J_i^F,i {\cal L}_i^{\rm F}]$ each $J^{\rm
  F}$ and ${\cal L}^{\rm F}$ splits into collinear and usoft parts in \SCETa,
$J^{\rm F} = T'(\omega_j){\overline{\cal J}}_{\omega_j} \Gamma {\cal H}_v$,
${\cal L}^{\rm F}=\overline {\cal Q}\, {\cal J} +\mbox{h.c.}$, where ${\cal
  J}$'s denote products of collinear fields. To factorize these time-ordered
products we follow Ref.~\cite{bps2}. From momentum conservation we have
$\omega_1+\omega_2 \to\bn\mcdot p_M$ of meson $M$, so we suppress this
dependence and let $\bar\omega=\omega_1\!-\!\omega_2$. With this notation we can
write
\begin{eqnarray} \label{TF}
 T_i^F &=&  T_i^\prime(\bar\omega) \int \!\! d^4 \!x \, 
  T\big[ \overline {\cal J}_{\!\bar\omega\,}\!(0)\Gamma {\cal H}(0) \: 
  \overline {\cal Q}(x) {\cal J}(x) \big] \\
 &=& T_i(\bar\omega) \int \!\! d^4 \!x \, 
  T\big[ \,\overline {\cal J}_{\!\bar\omega\,}\!(0) \Gamma_{\!c}{\cal J}(x) 
  \big] 
  T\big[\overline {\cal Q}(x)\Gamma_{\!s} {\cal H}(0)   \big]\,,\nn
\end{eqnarray}
where $T_i^\prime(\bar\omega)$ is $\{ C_\Gamma(\bn\mcdot p_M),
B_\Gamma^a(\bn\mcdot p_M), B_\Gamma^b(\bn\mcdot p_M,\bar\omega),
C_\Gamma(\bn\mcdot p_M) \}$. In the second line we performed a Fierz
transformation on the color and spin indices, absorbing prefactors to give
$T(\bar\omega)$, and dropping a $T^A\otimes T^A$ which gives no contribution in
\SCETb.  We now lower the off-shellness of the external collinear particles to
$p_c^2 \sim \LQCD^2$. The $T_i^F$ run exactly like their $J_i^F$ currents. Since
we have explicitly kept the usoft part of the momentum of collinear particles,
matching onto \SCETb amounts to setting $p_\perp^c = n \mcdot p^c = 0$ on
external lines and expanding the $T^F_{i}$'s.  Matching at $\mu_0\simeq
\sqrt{Q\Lambda}$ the usoft fields become soft (eg.~$Y \to S$), and the collinear
T-product matches onto a bilinear collinear quark operator in \SCETb,
\begin{eqnarray}
T\big[ \overline {\cal J}_{\bar\omega}(0) {\cal J}(x) \big] &=&
\delta(x^+)\delta^2(x^\perp) \int d\bar\eta \int\!\! dk^+ e^{\frac{i}{2}
k^+ x^-} 
\\
&& \!\!\!\!\times J(\bar\omega,\bar\eta,k^+)\:[\bar\xi_n^{\rm II} W \Gamma_{\!c}
\delta(\bar\eta\!-\!\bnP_+)W^\dagger \xi_n^{\rm II}]\,.\nn
\end{eqnarray}
The jet function
$J(\bar\omega,\bar\eta,k^+)$ is the Wilson coefficient for this matching
step. Inserting this in (\ref{TF}),
\begin{eqnarray}\label{TF2}
 T_i^F &=& \int\!\! d\bar\omega\, d\bar\eta\, dk^+\: T(\bar\omega) \:
   J(\bar\omega,\bar\eta,k^+)\: {\cal O}(\bar\eta,k^+)
   \,, \\
 && \hspace{-40pt}
 \mbox{\small ${\cal O}(\bar\eta,k^+) = [\,\bar\xi_n^{\rm II} W
        \delta(\bar\eta\!-\!\bnP_+)\Gamma_{\!c} W^\dagger \xi_n^{\rm II}\,]
  [\, \bar q^s S \Gamma_{\!s} \delta(\cP_+\!-\!k^+) S^\dagger h_v^s \,]$} ,\nn
\end{eqnarray}
where ${\cal O}(\bar\eta,k^+)$ is the full operator in \SCETb. Now taking the
\SCETb matrix element gives
\begin{eqnarray}\label{TFII}
  \langle M_n | {\cal O}(\bar\eta,k^+) | B_v\rangle = N f_M f_B \, 
     \phi_M(x) \phi_B^+(k^+) \,,
\end{eqnarray}
where $N$ is a normalization factor and $x=\bar\eta/(4E)+1/2$. Combining
Eqs.(\ref{TF2}) and (\ref{TFII}) reproduces Eq.~(\ref{fF}).

\begin{figure}[!t]
 \centerline{
  \mbox{\epsfxsize=3.3truecm \hbox{\epsfbox{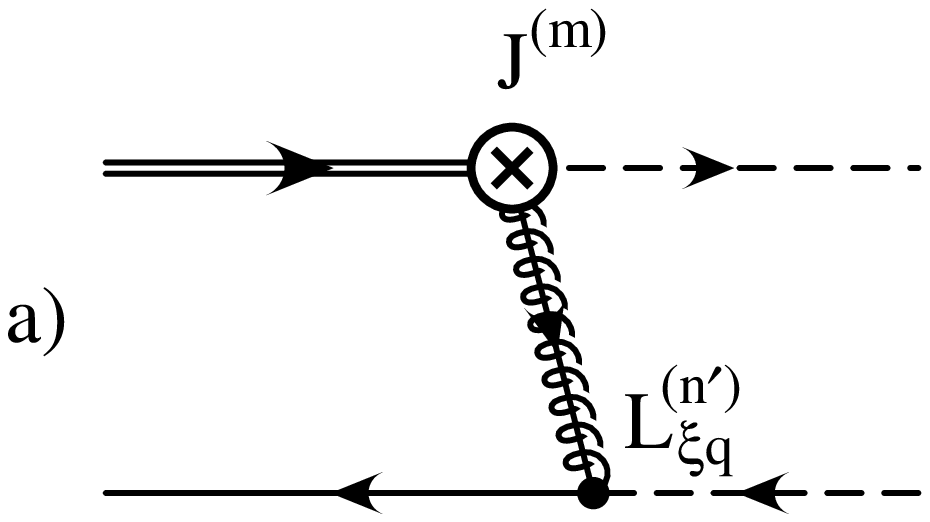}} }\hspace{0.4cm}
  \mbox{\epsfxsize=3.3truecm \hbox{\epsfbox{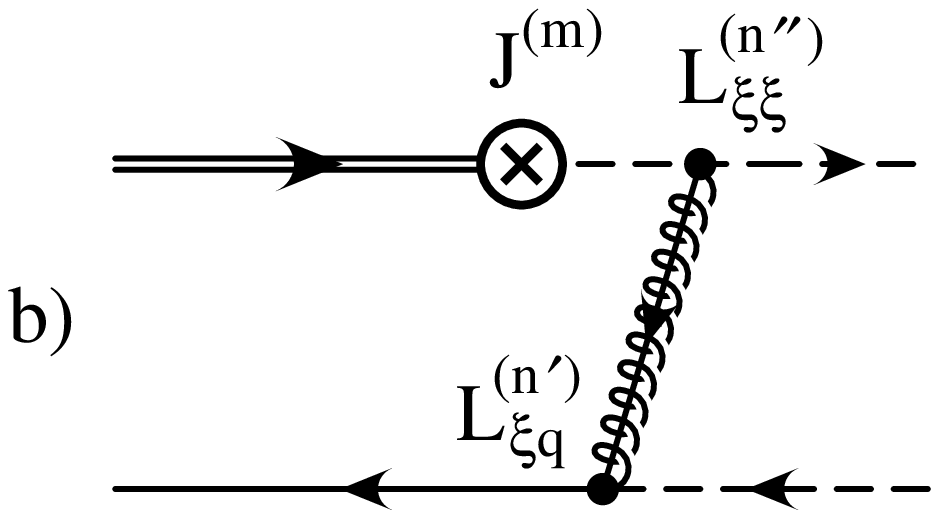}} }
  } 
\vskip-0.3cm
\caption[1]{Tree level graphs in \SCETa. The graphs in a) are from 
$T_{1,2,4}$, while those in b) are from $T_{0,1,3,4,5,6}$.}
\label{fig_eft} 
\vskip-0.5cm
\end{figure}
For the non-factorizable operators $T^{NF}_i$, it is not possible to write the
matrix elements as in $f^F$. Instead when matched onto \SCETb these terms give
$f^{NF}$ in Eq.~(\ref{fNF}) and should be understood to {\em define} the soft
nonperturbative effects for the form factors. It remains to show that they
satisfy the form factor relations~\cite{charles,bfps}.  Since the relevant
time-ordered products only contain the current $J_0$, the argument is the same
as in \cite{bfps}: any Dirac structure in heavy-to-light currents can be reduced
to only three, $\bar\xi_n W h_v$, $\bar\xi_n W \gamma^5 h_v$ and $\bar\xi_n W
\gamma_\perp^\mu h_v$. These three operators contribute only to $B \to P$, $B
\to V_{||}$ and $B \to V_\perp$, respectively, where $P$, ($V_{||}$, $V_\perp$)
denote pseudoscalar, (longitudinally, transversely) polarized vector mesons. For
$J_0$ this is true even in arbitrary time-ordered products with Lagrangian
insertions, since Lagrangians are parity even Lorentz scalars. The $f^F$ term
breaks these relations, but is calculable. At higher order in $\lambda$
non-factorizable contributions will also break these relations, since subleading
currents appear in time-ordered products with nonfactorizable Lagrangian
insertions.

The matrix elements of $T_{1,2}^F$ contain only $\phi_B^+$ to all orders in
$\alpha_s$ since inserting a projector next to $\xi_n$ in ${\cal L}_{\xi
  q}^{(1)}$, the $q_{us}$ appears as $\bar q_{us} {\nslash\bnslash}$ in the
Fierzed operators. On the other hand, $T_3^F$ (which may contribute at ${\cal
  O}(\alpha_s^2)$) has only $\bar q_{us} \bnslash$ and so is proportional to
$\phi_B^-$.  However, $T_3^F$'s matrix element involves $J_0$ and therefore
satisfies the same symmetry relations as the nonfactorizable matrix elements in
$f^{NF}$~\cite{ps1}. Therefore it can be absorbed into a redefinition of the
$\zeta_k^M$'s to all orders in perturbation theory.

The last step is to understand the power counting of the two contributions in
Eqs.~(\ref{fF},\ref{fNF}).  When we expand to match onto ${\rm SCET}_{\rm II}$
the new operators and coefficients scale with $1/Q$ in the same way as those in
${\rm SCET}_{\rm I}$, up to a global $1/Q$ from switching from the $\xi_n^{\rm
  I}$ to $\xi_n^{\rm II}$ fields. The one exception is $T_0^F$, since it is odd
in the number of $D_c^\perp$ derivatives and this extra $\perp$ gets suppressed
by at least one power of $\lambda$. Therefore, $T^F_i$ and $T^{NF}_i$ contribute
at the same order in $1/Q$ to the form factors. We find a generic form factor to
scale as $(\Lambda/Q)^{3/2}$, which is $\Lambda^2/Q^2$ suppressed compared to
the scaling in $m_b$ near $q^2_{\rm max}$ derived from HQET~\cite{HQET}.

We finally show that the endpoint singularities encountered in
(\ref{Wab}) do not occur in $f^F$ in the second step of matching. The
contributions of the time-ordered products at ${\cal O}(g^2)$ are shown in
Fig.~\ref{fig_eft}, and expanding we find $i{\cal A}_i= g^2\bar
\xi_n(p_1) X_i T^A h_v(p_b) \bar v(r) V_i T^A \xi_n(p_2)\, /P_g$ with
\begin{eqnarray}\label{Tnlo}
X_0\otimes V_0 &=& X_3\otimes V_3  = X_6\otimes V_6 = 0\nn\\
 X_1\otimes V_1 
 \!\!&=&\!\! 
  \frac{\gamma_\perp^\mu \bnslash \Gamma \otimes \gamma^\perp_\mu}{2\,\bn 
  \mcdot p}\,,   \quad
 X_2\otimes V_2
  = \frac{\Gamma {\nslash} \gamma_\perp^\mu \otimes 
 \gamma^\perp_\mu }{2m_b} \,,\nn\\
 X_4\otimes V_4
 \!\!&=&\!\! 
 \frac{1}{\bn \mcdot p_2} \Gamma\otimes \bigg[{\bnslash}{} -
  \frac{2 \Slash{p}_{2\perp}^{us}}{n \mcdot r}  \bigg], 
 \\
 X_5\otimes V_5
 \!\!&=&\!\! 
 \bigg[ \frac{\gamma_\perp^\mu\Slash{r}_\perp}
 {\bn \mcdot p \, n \mcdot r} + \frac{\Slash{p}_{1\perp}^{us}\gamma_\perp^\mu}
  {\bn \mcdot p_1 \, n \mcdot r} \bigg] \Gamma \otimes  \gamma^\perp_\mu \,.\nn
\end{eqnarray}
The $1/x^2$, $1/r_+^2$ singularities only exist in the non-factorizable $T_4$
and $T_5$, while the factorizable $T^F_{1,2}$ give non-singular jet functions.
This is not surprising, since in full QCD all endpoint singularities are
regulated by $\Lambda$. Thus, if (u)soft operators are properly included to
account for this region of momenta endpoint singularities will not arise.

As an example, for the form factor $f_+$ at leading order in $1/Q$ and all
orders in $\alpha_s$ we find 
\begin{eqnarray} \label{fplus}
 f_+ &\!\!\!=\!\!\!&  N_0 \int \!\! dx \, dz \, dr_+ \!\bigg[
 \frac{2E\!-\!m_B}{m_B} T_a(\mu_{\rm 0}) J_a(z,x,r_+,\mu_0,\mu)\: \nn\\
&&\!\!\!\!\! + \frac{2E}{m_b} T_b(z,\mu_{\rm 0}) J_b(z,x,r_+,\mu_0,\mu)\bigg]
 \phi_M(x,\mu)\: \phi_B^+(r_+,\mu)\nn\\[5pt]
&&\!\!\!\!\!  + C(Q,\mu) \, \zeta(Q,\mu)\,,
\end{eqnarray}
where $N_0=f_B f_\pi m_B/(4E^2)$ and the $Q$ dependence of $T_{a,b}$ and
$J_{a,b}$ is implicit. Here $T_{\{a,b\}}$ are the Wilson coefficients of the
currents $J^{(1a,1b)}$,  the jet functions $J_{a,b}$ are computed from the
$T^F_{1,2}$ time ordered products, and we have reabsorbed possible $\phi_B^-$
contributions from $T^F_3$ into $\zeta$. For the jet functions at order
$\alpha_s$ we find
\begin{eqnarray} \label{JaJb}
  J_a = J_b = \frac{\pi C_F}{N_c} \:
   \frac{ \alpha_s(\mu_{\rm 0})}{ x \, r_+}\: \delta(x-z) \,.
\end{eqnarray}
At tree level the coefficients satisfy $C = T_a = T_b = 1$ and using
Eq.~(\ref{JaJb}) the first term in Eq.~(\ref{fplus}) then agrees with the
non-singular hard contribution in~\cite{bf}. This simple approximation misses
double logarithms in $T_{a,b}(\mu_0)$ which may be larger than the single
logarithms resummed in the $\alpha_s(\mu_{\rm 0})$ for $\mu_0\simeq
\sqrt{Q\Lambda}$.  The one loop expression for $C(Q,\mu)$ can be found in
Eqs.~(33), (60) of \cite{bfps}. The non-perturbative matrix element $\zeta(E)$
is the reduced soft form factor describing decays to pseudoscalar mesons.

In this paper we proved a factorization formula for heavy-to-light decays
including spectator effects. The factorizable pieces are finite and determined
by one-dimensional convolutions. The nonfactorizable pieces include
non-perturbative gluon effects and satisfy form factor relations.  They are not
determined by the standard $k_\perp$ dependent light-cone meson wave functions,
which is different from the conclusion in~\cite{Kurimoto}. Our leading order
analysis needed the currents $J^{(1a,1b)}$, unlike the analysis in
Refs.~\cite{chay,bcdf} where these currents first enter at subleading order.

This work was supported in part by the DOE under grants DOE-FG03-97ER40546 and
DE-FG03-00ER-41132 and the NSF under grant PHY-9970781.


\begin{thebibliography}{99}
  
\bibitem{bl} For a review see S.~J.~Brodsky and G.~P.~Lepage, in {\it
    Perturbative Quantum Chromodynamics}, p.~93-240.


\bibitem{QCDsumsoft}
E.~Bagan {\it et al.},
Phys.\ Lett.\ B {\bf 417}, 154 (1998).

A.~Khodjamirian, {\it et al.},
Phys.\ Lett.\ B {\bf 410}, 275 (1997)

\bibitem{sing}
A.~Szczepaniak {\it et al.},
Phys.\ Lett.\ B {\bf 243}, 287 (1990);
B.~V.~Geshkenbein {\it et al.}, 
Phys.\ Lett.\ B {\bf 117}, 243 (1982);
M.~Beneke {\it et al.},
Nucl.\ Phys.\ B {\bf 606}, 245 (2001);
R.~Akhoury {\it et al.},
Phys.\ Rev.\ D {\bf 50}, 358 (1994).

\bibitem{bfps}
C.~W.~Bauer, S.~Fleming, D.~Pirjol and I.~W.~Stewart,
Phys.\ Rev.\ D {\bf 63}, 114020 (2001).

\bibitem{bfl}
C.~W.~Bauer, S.~Fleming and M.~Luke,
Phys.\ Rev.\ D {\bf 63}, 014006 (2001).

\bibitem{cbis}
C.~W.~Bauer and I.~W.~Stewart,
Phys.\ Lett.\ B {\bf 516}, 134 (2001).

\bibitem{bps2}
C.~W.~Bauer, D.~Pirjol and I.~W.~Stewart,
Phys.\ Rev.\ D {\bf 65}, 054022 (2002).

\bibitem{bps3}
C.~W.~Bauer, D.~Pirjol and I.~W.~Stewart,
Phys.\ Rev.\ D {\bf 66}, 054005 (2002).

\bibitem{bfprs}
C.~W.~Bauer, S.~Fleming, D.~Pirjol, I.~Z.~Rothstein, I.~W.~Stewart,
Phys.\ Rev.\ D {\bf 66}, 014017 (2002).

\bibitem{HQET} A.~V.~Manohar and M.~B.~Wise,
Cambridge Monogr.\ Part.\ Phys.\ Nucl.\ Phys.\ Cosmol.\  {\bf 10}, 1 (2000).

\bibitem{bf}
M.~Beneke and T.~Feldmann,
Nucl.\ Phys.\ B {\bf 592}, 3 (2001).

\bibitem{Kurimoto} T.~Kurimoto, H.~Li, A.~I.~Sanda,
Phys.\ Rev.\ D {\bf 65}, 014007 (2002).

\bibitem{Li}
H.~n.~Li,
Phys.\ Rev.\ D {\bf 66}, 094010 (2002).


\bibitem{NagashimaLi}
M.~Nagashima and H.~n.~Li,
arXiv:hep-ph/0210173.

\bibitem{grozin}A.~G.~Grozin and M.~Neubert,
Phys.\ Rev.\ D {\bf 55}, 272 (1997).

\bibitem{charles}
J.~Charles {\it et al.},
Phys.\ Rev.\ D {\bf 60}, 014001 (1999).

\bibitem{bpspc}
C.~W.~Bauer, D.~Pirjol and I.~W.~Stewart,
Phys.\ Rev.\ D {\bf 66}, 054005 (2002).

\bibitem{chay}
J.~Chay and C.~Kim,
Phys.\ Rev.\ D {\bf 65}, 114016 (2002).

\bibitem{rpi}
A.~V.~Manohar, T.~Mehen, D.~Pirjol, I.~W.~Stewart,
Phys.\ Lett.\ B {\bf 539}, 59 (2002).


\bibitem{bcdf}
M.~Beneke, A.~Chapovsky, M.~Diehl, T.~Feldmann, 
Nucl.\ Phys.\ B {\bf 643}, 431 (2002).

\bibitem{ps1}
D.~Pirjol and I.~W.~Stewart,
arXiv:hep-ph/0211251.







\end{thebibliography}
\end{document}